\documentstyle[11pt,aaspp4]{article}

\received{17 February 1997}
\accepted{13 November 1997}
\slugcomment{To be published in
the ApJ Suppl. Apr. 1998 Issue}

\newcommand{\cii}{[\ion{C}{2}]}
\newcommand{\kms}{km s$^{-1}$}
\newcommand{\vlsr}{$v_{\rm LSR}$}
\newcommand{\intensity}{erg s$^{-1}$ cm$^{-2}$ sr$^{-1}$}

\lefthead{Nakagawa et al.}
\righthead{\cii\ Line Survey of the Galactic Plane}

\begin{document}

\title{Far-Infrared \cii\ Line Survey Observations of \\
	the Galactic Plane}

\author{Takao Nakagawa, Yukari Yamashita Yui\altaffilmark{1}, 
	Yasuo Doi\altaffilmark{2}, Haruyuki Okuda,\\
	Hiroshi Shibai\altaffilmark{3}, and Kenji Mochizuki\altaffilmark{4}}
\affil{The Institute of Space and Astronautical Science,\\ 
	Yoshinodai 3-1-1, Sagamihara, Kanagawa 229, Japan}

\and

\author{Tetsuo Nishimura\altaffilmark{5} and Frank J. Low}
\affil{Steward Observatory, University of Arizona, Tucson, AZ 85721}

\altaffiltext{1}{Present Address: Communications Research Laboratory,
	Nukui-kitamachi 4-2-1, Koganei, Tokyo 184, Japan}
\altaffiltext{2}{Present Address: Department of Earth Science and Astronomy, 
	University of Tokyo, 
	Komaba 3-8-1, Meguro-ku, Tokyo 153, Japan}
\altaffiltext{3}{Present Address: Department of Astrophysics,
	Nagoya University,
	Furo-cho, Chikusa-ku, Nagoya 464-01, Japan}
\altaffiltext{4}{Present Address: Department of Astronomy, University
	of Texas at Austin, Austin, TX 78712-1083}  
\altaffiltext{5}{Present Address: National Astronomical Observatory,
	Osawa 2-21-1, Mitaka, Tokyo 181, Japan}

% ---------------------------------------------------------------

\begin{abstract}
We present results of our survey observations
of the \cii\ 158 \micron\ line emission from the Galactic
plane using the Balloon-borne Infrared Carbon Explorer (BICE).
Our survey covers a wide area 
(350\arcdeg $\lesssim l \lesssim$ 25\arcdeg, 
$|b| \lesssim$ 3\arcdeg) with a spatial
resolution of 15\arcmin. 
We employed a new observing method called 
the ``fast spectral scanning''
to make large-scale observations efficiently.

Strong \cii\ line emission was detected from 
almost all areas we observed.
In the general Galactic plane,
the spatial distribution of the \cii\ line emission
correlates very well with that of
far-infrared continuum emission, but
diffuse components are more prominent in the  \cii\ line emission;
the $I_{\rm [CII]} / I_{\rm FIR}$ ratio is
$\sim$ 0.6 \% for diffuse components but is $\sim$ 0.2 \%
for compact sources such as active star-forming regions.
In the Galactic center region, on the other hand, 
the distribution 
of the \cii\ line emission is quite different from that of
the far-infrared continuum emission, and the $I_{\rm [CII]} /
I_{\rm FIR}$ ratio is systematically lower there.

The FWHM velocity resolution of our instrument is 175 \kms, but 
we determined the central velocity of the line at each observed point
very precisely with statistical errors as small as $\pm$6 \kms.
The longitudinal
distribution of the central velocity clearly shows the differential
rotation pattern of the Galactic disk and also violent velocity 
fields around the Galactic center. 

\end{abstract}


% ---------------------------------------------------------------
\section{INTRODUCTION}

The far-infrared \cii\ fine structure line ($^{2}P_{3/2} \rightarrow$
$^{2}P_{1/2}$, 157.7409 \micron,
\cite{coo86}) has been predicted
to be the dominant coolant of neutral interstellar
gas (e.g., \cite{dal72}), 
and is expected to determine
the cooling rate (and consequently the temperature) 
of general interstellar clouds.
This is because 
(1) the C atom is easily ionized by general
interstellar radiation field, since its ionization energy (11.3 eV)
is lower than that of H atom (13.6 eV), and
(2) the C$^{+}$ ion is easily
excited collisionally  from its ground level
($^{2}P_{1/2}$) to the first fine-structure level ($^{2}P_{3/2}$)
due to its relatively small energy difference ($\Delta E/k = 91$ K) 
and to its small critical density ($n_{\rm cr} = 3 \times 10^3$ cm$^{-3}$
for \ion{H}{1}).
Hence the line is expected to be a useful probe
of the energy budget of general interstellar gas.
Moreover, since the \cii\ line is very bright,
its observations can trace 
general interstellar clouds on a large
scale, and hence can reveal the global structure of the Galaxy. 

The \cii\ line cannot be observed from the ground because of
heavy absorption due to atmospheric water vapor.
Thus most of previous observations have been made
from the Kuiper Airborne Observatory (KAO) 
(see, e.g., \cite{haa95})
or from balloon-borne telescopes (e.g., \cite{mat89}, \cite{shi91}, 
and \cite{miz94}).
Most of these observations were
restricted to small areas around active star-forming regions
or to active regions in starburst galaxies.
Since, in these regions, interstellar clouds are directly exposed to 
very strong ultraviolet radiation from young OB stars,
photodissociation region models
with rather intense radiation
field and high density gas (e.g., \cite{tie85})
have been successfully applied to interpret the observations.
 
The clouds in these active regions, however, do not represent
general clouds in the Galaxy, and most of interstellar clouds 
in the Galaxy have lower gas density and are exposed to much weaker 
radiation field. 
Hence large-scale \cii\ line survey observations of
various kinds of interstellar clouds
are required to reveal general characteristics of
interstellar clouds in the Galaxy.
Moreover, since the total emission from ordinary galaxies
is dominated by emission from these general clouds,
large-scale \cii\ line observations of the Galaxy are 
also important to our understanding of 
interstellar clouds in other galaxies. 

Few observations of large-scale \cii\ line emission 
from the Galaxy have been made so far.
The pioneering observations by Stacey et al. (1985) using 
an air-borne telescope 
and those by Shibai et al. (1991) using a balloon-borne telescope
revealed that the \cii\ emission
is bright and ubiquitous, but the
areas of these observations were very limited.
The Far-Infrared Absolute Spectrophotometer
(FIRAS) aboard the Cosmic Background Explorer (COBE) 
made the first all-sky observation of the far-infrared \cii\ line 
(\cite{wri91}; \cite{ben94}), and revealed
that, on a Galactic scale,  
the \cii\ line is the brightest emission
line in the far-infrared and  sub-mm wavelengths. 
However, the beam size of FIRAS was too large
(7\arcdeg\ FWHM) to resolve detailed structures of
the \cii\ line emission from the Galactic plane,
and it is difficult to compare the FIRAS data directly
with other surveys (radio continuum, IRAS, \ion{H}{1},
and CO) of the Galactic plane.
Hence, in order to study general interstellar clouds,
it has been necessary to make  new \cii\ line  observations, 
which cover a significant portion
of the Galactic plane with a spatial resolution
comparable to those of other Galactic plane surveys 
made in other wavelengths.

We have made large-scale survey observations
of the \cii\ 158 \micron\ line
emission from the Galactic plane using a balloon-borne telescope. 
Our goal is to map the entire inner Galactic plane
with a sufficiently good spatial resolution ($15\arcmin$)
to reveal detailed structures of the Galactic plane
and can be compared directly with other Galactic plane surveys.
Preliminary results of our \cii\ survey were
discussed by Okuda et al. (1994) and by Nakagawa et al. (1995a).
This paper fully describes our instrumentation, observations,
and the data analysis, and presents the 
first results of our Galactic survey observations,
which covers 350\arcdeg $\lesssim l
\lesssim$ 25\arcdeg\ and  $|b| \lesssim$ 3\arcdeg.
The main characteristics of the large-scale \cii\ emission
are also discussed briefly, but
detailed interpretation of the large-scale \cii\ line emission
will be discussed in a later paper.

Recently, the Far-Infrared Line Mapper (FILM)
on the Infrared Telescope in Space (IRTS)
also made large-scale observations of the \cii\ line 
(\cite{shi96}, \cite{mak96})
with the spatial resolution ($8\arcmin\ \times
13\arcmin$) similar to that of the present 
observations. FILM has better sensitivity than that of the present work, 
but, as for the Galactic plane, covers only limited areas (\cite{mur96}).
Hence the current observation is more suitable for the systematic study of 
strong emission from the Galactic plane while 
the FILM survey can reveal much weaker emission at high Galactic
latitudes. These two sets of observations are thus complementary.

% ---------------------------------------------------------------
\section{INSTRUMENTATION AND OBSERVATIONS}

\subsection{Balloon-borne Infrared Carbon Explorer (BICE)}
We observed the far-infrared \cii\ line emission from the Galactic
plane using the Balloon-borne Infrared Carbon Explorer (BICE),  
which is customized for large-scale, 
far-infrared spectroscopic observations.

The sensitivity of far-infrared observations with
air-borne or balloon-borne telescopes
are generally limited by photon fluctuations 
of atmospheric and instrumental radiation (foreground radiation).
The atmospheric radiation is still very strong
at airplane altitudes. 
At balloon altitudes, on the other hand,
the atmospheric radiation is significantly reduced (\cite{tra76}),
and the instrumental radiation generally dominates the foreground 
radiation.
Hence, in order to achieve a good sensitivity by taking advantage of 
good atmospheric condition at balloon altitudes,
it is essential to reduce the instrumental radiation.

When a conventional telescope at ambient temperature is used,
the radiation from the telescope itself generally dominates 
instrumental radiation.
The BICE telescope is also at ambient temperature, and, 
in order to reduce the instrumental radiation,
we suppressed the emissivity of the telescope in two ways 
(see also \cite{nak93}).
First, we employed oversized optics.
The physical size of the primary mirror
is 35 cm, but the effective size, 
which is determined by a cold Lyot-stop in the 
spectrometer, is only 20 cm, and the outer part of the mirror is
used to minimize the spill-over of radiation due to diffraction.
Second, we employed an offset
optics design, which removed the secondary mirror and its support
from the optical path.
The combination of the offset and oversized optics dramatically
reduced the radiation from the telescope,
and the emissivity of the telescope itself
was measured to be only about 1 \% at 158 \micron.

The BICE focal
plane instrument is a tandem Fabry-Perot spectrometer, 
which consists of two Fabry-Perot interferometers:
one is a high-order interferometer to scan the wavelength
(Scanning Fabry-Perot, SFP), 
and the other is a low-order interferometer 
(Fixed Fabry-Perot, FFP) as an order-sorter of the SFP.
The velocity resolution 
($\Delta v$) is 175 \kms\ (FWHM).
Both of the interferometers, together with other optics of the
spectrometer and a far-infrared detector, 
are cooled to 2 K with liquid helium,
and the instrumental radiation from the spectrometer
is completely negligible. 
The detector is a stressed Ge:Ga photoconductor and 
is described by Hiromoto et al. (1989).

When combined with the BICE telescope,
the spectrometer was designed to have a beam size of 
12\arcmin\ (FWHM). Figure 1 shows the beam pattern
obtained from laboratory measurements.
We measured the beam pattern twice: once before the observations
and once after the observations. The two measurements were consistent. 
The beam pattern can be fitted
by a Gaussian with a size of
$12\farcm4$ (FWHM) and an effective solid angle of 
$1.5 \times 10^{-5}$ sr. 

\subsection{Fast Spectral Scanning Method}

The BICE experiment is dedicated for 
large-scale survey observations 
of \cii\ line emission.
However, the spatial chopping method, which has been used 
for many infrared observations,
is not suitable for large-scale observations, since
signals from a reference beam could be contaminated by 
residual, extended emission (so-called ``self-chopping'').
Hence, instead of the conventional spatial chopping method,
we employed a ``fast spectral scanning
method'' to cancel the foreground radiation.

In the ``fast spectral scanning method'', the signal is modulated
not in the spatial domain but in the spectral domain.
The modulation was achieved by continuously sweeping the SFP
back and forth at 2.7 Hz in the velocity
range of $\Delta v_{\rm SCAN} = 520$ \kms\ around the expected wavelength
of the \cii\ line.
We determine the line intensity from the modulated signal.
Moreover, from the phase of the modulation, we also obtain the
central velocity of the line.
Details of the data analysis are discussed in $\S$ 3.2.
Since this method does not require nearby reference positions,
it is suitable for mapping observations
of spatially extended line emission.

\subsection{Attitude Control}

The BICE telescope is mounted on an alt-azimuth pointing system,
and only the azimuthal axis is servo-controlled by driving a
reaction bar. 
Observations were made by rotating the whole gondola around the
azimuthal axis back and forth at a fixed elevation.
With the diurnal motion of the sky, we were able to make a two-dimensional 
map of the Galactic plane.
The azimuthal scan was made
at a speed of 12\arcmin\ s$^{-1}$ with a scan width of 8\arcdeg.
The spectral scanning was simultaneously made at 2.7 Hz as mentioned before,
and thereby one spectral profile was obtained at about every one third 
of the beam width.
We moved the elevation axis intermittently
to change observing areas.

A geomagnetic aspect sensor was
used as a reference of the azimuthal control. 
The stability of the attitude control was about 1\arcmin\ r.m.s.
when external perturbations were not severe.
Since the azimuthal angle determined from the geomagnetic sensor 
has absolute errors as large as 1\arcdeg,
we used a visible star sensor to determine absolute positions.
The uncertainty of the final position reconstruction is 
about 5\arcmin\ r.m.s.

\subsection{Observations}

We made two balloon flights from the National Scientific
Balloon Facility in Palestine, Texas, USA in 1991: 
one on May 26 and the other on June 12. 
In the first flight, the total emissivity 
which included both instrumental and atmospheric radiation, was only 3 \%,
but, due to some excess noise (probably electric interference), 
the quality of the data was not good.
In the second flight, the emissivity was
rather high (6 \%), since the mirror had become somewhat
dirty during the recovery after the first flight.
However, the quality of the data was
much better, and hence we concentrate 
on the data from the second flight in the following discussion.
The system NEP during the second flight was 
$6 \times 10^{-16}$ W Hz$^{-1/2}$. 
Observation parameters are summarized in Table 1.

In the second flight, the floating altitude was
37 - 38 km and the float duration was about 8 hours. 
The total observation
time for the Galactic plane was about 6 hours.
During the same flight, we also observed 
the $\rho$ Ophiuchi dark cloud (\cite{yui93})
and the Cygnus-X region (\cite{doi93}).
In this paper, we concentrate on the Galactic plane data.

% ---------------------------------------------------------------
\section{DATA ANALYSIS}

\subsection{Foreground Radiation}

We obtained more than 50,000 spectral profiles of the Galactic plane.
Each spectral profile contains 
not only astronomical emission but also foreground radiation,
which is the combination of instrumental 
and atmospheric radiation.
In order to remove the foreground radiation
and to extract astronomical emission, 
we assume that, at high Galactic latitudes, 
the astronomical emission is negligible
and observed profiles contain only the foreground radiation.
Figure 2 shows the observed area together with the foreground areas 
where we assume that astronomical emission is negligible.
For comparison, we also show a far-infrared continuum emission
($I_{\rm FIR}$) map,
which is calculated from the IRAS 60 \micron\ and 100 \micron\
maps (\cite{ira86})
as follows.
\begin{equation}
\left(\frac{I_{\rm FIR}}{\rm erg\ s^{-1}\ cm^{-2}\ sr^{-1}}\right) =
3.25 \times 10^{-11}
\left(\frac{f_{\nu}(60 \micron)}{\rm Jy\ sr^{-1}}\right)
+ 1.26 \times 10^{-11}
\left(\frac{f_{\nu}(100 \micron)}{\rm Jy\ sr^{-1}}\right)
\end{equation}
This relation was originally defined for point 
sources (\cite{hel88}), and we used the definition
also for extended sources to represent the radiation
between 40 and 120 \micron.
The spatial resolution is smoothed to 15\arcmin.
We took the foreground areas basically as $|b| > $3\arcdeg,
but changed the areas slightly
when the scan width is not enough (e.g., at $l \approx -5$ \arcdeg) 
or when astronomical emission extends to high latitudes
(at $l \approx 19$ \arcdeg).

The average of astronomical far-infrared emission ($I_{\rm FIR}$) 
in the foreground areas
is about $3 \times 10^{-3}$ \intensity,
which corresponds to \cii\ line emission ($I_{\rm [CII]}$) of
$2 \times 10^{-5}$ \intensity\
(Here we assume $I_{\rm [CII]} / I_{\rm FIR} = 6 \times 10^{-3}$; see \S4.2).
We also get $I_{\rm [CII]}$ = $2.3 \times 10^{-5}$ \intensity\
at $|b| = 3.5$\arcdeg\ by extrapolating the FIRAS measured 
intensity cosecant of the high
Galactic latitude ($|b| > 15$\arcdeg) \cii\ line emission (\cite{ben94}).
In summary, our observations have a negative offset of 
$I_{\rm [CII]} \sim 2 \times 10^{-5}$ \intensity\,
which is comparable with our 3$\sigma$ detection limit 
($I_{\rm [CII]} = 2 \times 10^{-5}$ \intensity; see \S3.3).
Since we concentrate on the \cii\ emission from the Galactic plane
with $|b| \leq $ 3 \arcdeg, we make no correction for this offset
in most of the following discussion. This offset, however, should be taken
into account, when our observations are compared with 
larger scale observations (\S4.2).

\subsection{[C II] Line Emission}

Each observed spectral profile was corrected for foreground radiation
as described above.  
Since the instrumental spectral profile is much wider
than intrinsic astronomical line profiles, observed
spectral profiles corrected for foreground radiation 
are dominated by the instrumental profile.
We hence fitted each observed line profile 
with a single Lorentzian and a linear base line.
The Lorentzian profile is a good approximation for the instrumental profile.
However, the observed profile is not exactly the instrumental profile;
the astronomical line emission
modifies the observed profiles from the instrumental profile
by slightly broadening their widths
and by shifting their central positions.
Taking these effects into consideration,
we let the central position, the width, and the height of a Lorentzian
be free parameters for the regions with strong \cii\ line emission
(typically $I_{\rm [CII]} \geq 6 \times 10^{-5}$ \intensity),
where we can determine the three parameters accurately.
We then calculated the \cii\ line intensity from the fitted width and height, 
and derived the central velocity 
from the fitted central position of the Lorentzian.

On the other hand,
for the regions with weak \cii\ line emission
(typically $I_{\rm [CII]} < 6 \times 10^{-5}$ \intensity),
it is rather difficult to determine all of the three parameters
of the Lorentzian independently.
We hence assume that, in these regions,
the \cii\ line emission is dominated by radiation from
the general Galactic plane.
We can thus expect the central position
and width of the line to be almost constant along a strip across the
Galactic plane with the constant Galactic longitude.
Figure 3 shows that each spatial scan is across the Galactic plane with almost 
constant longitude.
Hence, in the regions with weak \cii\ line emission,
we assumed that the width and the central position
was constant in each spatial scan and changed gradually from scan to scan.
In each spatial scan, we calculated the mean central position and 
line width, both of which were weighted by line intensities,
from the fitted results in the regions with strong
\cii\ line emission.
Thus, for the regions with weak \cii\ line emission,
we fixed the width and the central position
to the mean values in each spatial scan  
and fitted the height.
From the fitted height and the assumed width,
we calculated the line intensity.

Figure 4 shows an example of this fitting procedure for one spatial scan.
This scan started at $(l, b) = (-4\fdg5, -2\fdg8)$
and ended at $(l, b) = (-11\fdg5, 3\fdg6)$.
The horizontal axis is the Galactic latitude.
We show line intensities in Figure 4a, central positions
in Figure 4b, and line widths in Figure 4c.
In the region (A), where the \cii\ line intensity is strong, 
all of the width, the velocity, and the intensity of the line
were derived independently at each observed point. 
In the regions (B), where the \cii\ line intensity is weak, 
only the the line intensity were derived 
independently at each observed point.
The regions (C) correspond to the foreground areas.
Flat lines in the regions (B) and (C) in Figures 4b and 4c indicate that 
both of the width and the velocity of the line were assumed
to be constant in one scan.

The line intensities derived thereby were binned spatially into 
3\arcmin\ grids to make an intensity map.
The data for grids without observations were
linearly interpolated from the data in nearby bins.
The map was then smoothed by a circular Gaussian with a FWHM
of 8.4\arcmin\ to remove artificial strip-like features
due to the spatial scan.
The final spatial resolution was broadened to 15\arcmin\ 
by this smoothing.

\subsection{Intensity Calibration}

We observed M~17
as a \cii\ line flux calibrator during the same flight. 
Matsuhara et al. (1989) obtained a two-dimensional map 
of the \cii\ line emission from M~17.
We calculated the \cii\ line flux within our beam size from their map,
and corrected the result for  residual \cii\ line emission 
in the regions where they determined their baseline.
The estimated \cii\ line flux within our beam at the peak of M~17
is  $1.4 \times 10^{-8}$ ergs s$^{-1}$ cm$^{-2}$. 
The final calibration uncertainty of the absolute intensity
is about $\pm$ 35 \%.

We analyzed two data sets independently; one data set was obtained
when the SFP scanned from shorter to longer wavelengths
(SFP-forth), and the other contains scans in the opposite manner (SFP-back).
From the two independent maps thus derived, we can estimate the reliability
of our data.
We calculated the difference of intensity at each pixel between the
two maps. Figure 5 shows the distribution of the difference.
The distribution is very close to a Gaussian, and
the systematic offset between the two maps are quite small
($\Delta I_{\rm [CII]} < 1 \times 10^{-6}$ \intensity),
which indicate that 
the noise is dominated by statistical fluctuations and systematic
differences between the two maps are quite small.
We obtained the final map by averaging the two independent data maps
(\S4.1), and we can also estimate the noise level of the final map from Figure 5.
Since Figure 5 is the distribution of flux differences 
between the two maps and is dominated by statistical fluctuations,
the noise level of each map is $1/\sqrt{2}$ of that of Figure 5, 
and the noise level of the final averaged map is 1/2 of that of Figure 5.
The detection limit of the \cii\ line derived thereby is
$2 \times 10^{-5}$ \intensity\ (3 $\sigma$).

The uncertainty of the final intensity map is affected
not only by this statistical noise but
also by the uneven scanning paths
of our observations. 
Figure 3 shows that, 
for most of the regions along the Galactic ridge,
the step between adjacent scanning paths
is smaller than a half of the beam size,
which means the data were fully sampled there.
However, in some regions at high latitudes
(and also at a strip across the Galactic plane at $l \approx$ 23\arcdeg),
the distance between adjacent scans
is as large as the beam size.
In these regions, the data were not fully-sampled,
and peak intensities and positions 
of compact sources are relatively unreliable.
The intensity of diffuse components is reliable to the statistical limit
even in these regions.

\subsection{Velocity Calibration}

We used an atmospheric line 
(O$_{3}$, $20_{8\ 12} \rightarrow
19_{7\ 13}$, 157.6121 \micron; \cite{rot87}) 
and astronomical \cii\ line emission
to calibrate the observed wavelength,
and used M~17 and NGC~6334
as astronomical \cii\ line calibration sources.
The \cii\ line velocity of these sources were
adopted from radio observations of carbon recombination lines;
\vlsr\ = 15.0 \kms\ for M~17 (\cite{pan77}) and \vlsr\ = $-3.3$ \kms\
for NGC~6334 (\cite{mcg81}). The observed central wavelength 
of the \cii\ line was 
converted to \vlsr\ following Gordon (1976).   
The uncertainty of the absolute velocity calibration is
$\pm$10 \kms.

We made velocity maps by binning the data spatially 
into 12\arcmin\ grids. We applied no smoothing for the velocity data.
We also analyzed two data sets (SFP-forth and SFP-back) 
independently, and Figure 6 shows the distribution of velocity difference
at each pixel between the two data sets.
The distribution is close to a Gaussian,
and the systematic offset between the two maps 
($\Delta$ \vlsr\ $\approx 3$ \kms)
is within the uncertainty
of the absolute velocity calibration.
The final velocity map is the average of the two independent maps,
and Figure 6 shows that the statistical uncertainty of 
the final velocity map is $\pm$6 \kms\ (1 $\sigma$).

% ---------------------------------------------------------------
\section{RESULTS}

\subsection{General Characteristics of [C II] Line Emission}

Figure 7a shows a velocity-integrated \cii\ line intensity map
of the Galactic plane by false colors, and Figure 8 shows the
same data in the form of a contour map. 
These maps were obtained by averaging the SFP-forth and SFP-back data 
sets as described in $\S$ 3.2. 
Figure 8 also shows the observed area by shade and
identifications of bright sources.

As is shown in Figures 7a and 8, 
we have detected strong \cii\ line emission from most of the area we observed.
We see strong diffuse emission as well as
many compact, bright sources.

Table 2 is a list of bright \cii\ line 
($I_{\rm [CII]} \geq 3 \times 10^{-4}$ \intensity) peaks. 
Column 1 is a sequential number.
Columns 2 and 3 give the source peak position
in Galactic coordinates. 
%The uncertainty of the peak position
%is mostly due the uncertainty of the absolute position determination
%($\pm 5\arcmin$, see $\S$ 2.3) and partly due to the uneven sampling
%of the data (Fig.3).  
Column 4 shows the observed peak intensity.
Columns 5, 6, and 7 are only for discrete sources.
We assume that a \cii\ line emission feature is a discrete source
if the feature stood out clearly from the immediate base level
with our spatial resolution.
Column 5 show the base level intensity,
which is the average intensity of diffuse emission around the source.
Columns 6 and 7 list source sizes 
in Galactic longitude and latitude, respectively.
The source sizes are defined as FWHM sizes of the
foreground-subtracted source signals. 
Since the spatial distribution of the \cii\ emission is rather
complicated, it is difficult to define the base level intensities
and the source sizes precisely. Hence they have large uncertainties,
and should be treated accordingly.
Column 8 gives the observed source velocity at the peak
of the source (see also \S4.3).
Column 9 shows the source identification when it is possible.

Most of bright, compact sources correspond
to young \ion{H}{2} regions and active star-forming regions
(Table 2).
Tielens \& Hollenbach (1985) defined photodissociation regions (PDRs)
as the regions where far-ultraviolet (FUV) radiation dominates the
heating of interstellar matter and
determines the chemical composition of gas.
Around young OB stars, dense molecular clouds are irradiated by intense
radiation fields from the hot stars, and thus, warm, dense PDRs are formed.
Both strong radio thermal continuum emission (e.g., \cite{han87}),
which traces UV radiation from young OB stars,
and strong CO emission (e.g., \cite{dam87}),
which traces molecular clouds,
are associated with
most of the bright, compact \cii\ sources. 
Hence the \cii\ line emission from these sources
is attributed to the warm, dense PDRs.

On the other hand,
there are some \ion{H}{2} regions
with strong \cii\ line emission but with weak CO line emission.
The most notable example is S 54 at $(l, b) 
\approx (18\fdg5, 2\arcdeg)$.
Since this is a large, evolved  
\ion{H}{2} region (\cite{joh82}, \cite{mul87}),
we suppose that most of molecular clouds there have
been photodissociated significantly.
If this is the case, we can explain the strong \cii\ emission
with weak CO emission. We call this type of cloud a 
translucent PDR (see also \cite{van88}; \cite{doi93}; \cite{nak95b}).

Figure 7a also shows strong diffuse \cii\ line emission,
which is not directly associated with compact sources.
The most promising energy source of this emission is
diffuse interstellar UV radiation.
Shibai et al. (1991) suggested that the diffuse \cii\ emission comes from
PDRs illuminated by general interstellar UV radiation.
On the other hand, Heiles (1994) suggested that the extended
low-density warm ionized medium (ELDWIM) was the most important
candidate for the global source of the diffuse \cii\ line emission.
Petuchowski \& Bennett (1993) argued that both of them contributed to the 
diffuse \cii\ line emission.
The origin of the diffuse \cii\ line emission will be discussed 
in a later paper.

\subsection{Latitudinal Profile}

Figure 9 shows the latitudinal profile of the \cii\ line emission
together with those of \ion{H}{1} integrated line intensity (\cite{har97}), 
far-infrared continuum intensity (IRAS), and
CO 1-0 integrated line intensity (\cite{dam87}). 
We calculated the profiles by averaging
observed signals at 5\arcdeg $\leq l \leq $25\arcdeg.  
In order to get the profiles for the general Galactic plane,
we excluded the Galactic center region for this estimate.
We obtained far-infrared continuum intensity
using equation (1), and smoothed its spatial resolution
to that of the \cii\ line emission (15\arcmin). 
On the other hand, bin sizes of 
\ion{H}{1} and CO profiles are both 30\arcmin.
The peak intensity of each profile is normalized to unity.

All of the profiles have their peaks around $l \sim$ 0\arcdeg,
but show different widths.
The profiles are not simple Gaussian distributions.
For simplicity, we use the FWHM measured from the base line
which connects the signal at $l$ = -4\arcdeg\ and that at $l$ = 4\arcdeg.
This FWHM thus shows the width of the latitudinal profile of a 
component at $|b| \leq$ 4\arcdeg.

The latitudinal profile of the \cii\ line is 
similar to that of the far-infrared continuum,
but the FWHM of the \cii\ line (1\fdg32) is slightly wider than
that of far-infrared continuum (1\fdg18).
Both of them are wider than that of the CO line (1\fdg07),
and are much narrower than that of \ion{H}{1} (1\fdg96).
All of the above widths are observed values and are not corrected for
beam sizes.  

Shibai et al. (1991) also obtained latitudinal profiles
of the \cii\ line and other tracers of interstellar matter
for one cross-cut at $l \sim$ 31\arcdeg\ across a giant \ion{H}{2}
region W43. Since their observation covered a narrower latitude range
($|b| <$ 1\fdg2) than ours,
their observed \cii\ profile shows a narrower width 
(FWHM $\sim$ 0\fdg7, W43 excluded), and is consistent with
our observed profile.

Stacey et al. (1985) measured latitudinal profiles 
of the \cii\ line emission along three cross-cuts (at $l$ = 2\fdg16,
7\fdg28, and 7\fdg80) across the Galactic plane. 
Their observed peak intensity at $l$ = 7\fdg28
(1.1 $\times 10^{-3}$ \intensity) 
is about four times as large as our measured peak intensity
(2.7 $\times 10^{-4}$ \intensity) at $l$ = 7\fdg30. 
Petuchowski et al. (1996) also measured the \cii\ distribution
at $l$ = 7\fdg28, assuming the centroid of the distribution
determined by Stacey et al. (1985). 
Their observed peak intensity (3.7 $\times 10^{-4}$ \intensity) 
is consistent with our measured intensity at $l$ = 7\fdg30
within absolute calibration uncertainties. 
The FWHM of the distribution (43\farcm8) measured by  
Petuchowski et al. (1996) is also consistent with our value (41\farcm2),
but their assumed centroid of the \cii\ distribution
($b$ =  -$15\farcm1$) is clearly shifted from our observed
peak position ($b$ =  -6\arcmin). Since we measured the \cii\ 
line distribution more directly,
we believe that our measurement is more reliable than 
these previous measurements.

In order to obtain real scale heights, we
need to know radial distribution of each emission
as well as observed latitudinal distribution, and hence we have to be
careful to interpret the observed latitudinal distribution alone.
But the small FWHM of the \cii\ line emission and
its good spatial correlation with far-infrared continuum suggest
that radiation from young stars is responsible for a significant
fraction of the \cii\ line emission. 

\subsection{Longitudinal Distribution}

Figure 10 shows longitudinal distribution of the \cii\ line
emission observed by BICE together with the result
by FIRAS (\cite{ben94}). 
We calculated the BICE \cii\ distribution by integrating the 
observed line intensity over $|b| \leq$ 3\arcdeg\ at each longitude. 
We also added an uniform
offset of 2.3 $\times 10^{-5}$ \intensity (see  $\S$ 3.1)
to the BICE result, in order to compare the distribution 
with that of FIRAS. 
The FIRAS result is integrated over $|b| \leq$ 5\arcdeg\ (\cite{ben94}).

General profiles of the observed \cii\ line emission by BICE
is consistent with those of FIRAS, except for the region
near the Galactic center;
the \cii\ peak is offset toward negative longitude
by a half of the bin size in the FIRAS result.
But this offset is much smaller than the beam size of FIRAS (7\arcdeg),
and the pixel quantization error is likely to be responsible
for this offset.

One more thing we should note is that the relative intensity
between the two measurements. The total flux measured by  
BICE at $|b| \leq$ 3\arcdeg\ and 350\arcdeg $\leq l \leq$ 25\arcdeg\ is 
$6.0 \times 10^{-6}$ erg s$^{-1}$ cm$^{-2}$, while
that of FIRAS at $|b| \leq$ 5\arcdeg\ and the same longitudinal range
is $1.0 \times 10^{-5}$ erg s$^{-1}$ cm$^{-2}$.
Although we cannot draw a definitive conclusion here, 
this result suggests
that BICE observations may show systematically 
smaller ($\sim$ 65 \%) fluxes than FIRAS observations.
It is, however, very difficult to compare the two observations directly,
since (1) the effective solid angle of the both observations
differ by almost three orders of magnitude, (2) we do not
know the exact offset for the BICE measurement,
and (3) the distribution of the \cii\ line emission at 
3\arcdeg $< |b| <$ 5\arcdeg\ is uncertain.

\subsection{[CII] Line Emission and Far-Infrared Continuum Emission}

Since the \cii\ line is the dominant
coolant of the general interstellar gas, the \cii\
line intensity ($I_{\rm [CII]}$)
traces the heating rate of interstellar gas.
The far-infrared continuum emission
($I_{\rm FIR}$), on the other hand,
is a good measure of the total luminosity.
Hence the $I_{\rm [CII]}/I_{\rm FIR}$ ratio traces
the ratio of gas heating rate to the total energy input.

In this subsection, we briefly discuss the correlation between 
the \cii\ line emission and far-infrared continuum emission.
We calculated $I_{\rm FIR}$ using equation (1), and
in order to compare far-infrared continuum data directly
with our \cii\ data and also to
remove zodiacal emission from the continuum data,
we subtracted a certain offset from the data.
We determined the offset at each longitude
by linearly interpolating two data sets: one averaged over
$3\arcdeg < b \leq$ 4\arcdeg\ and the other 
over $-4\arcdeg \leq b < -3\arcdeg$. 
These regions roughly correspond to the foreground
areas in Figure 2.
Figure 7b shows this corrected far-infrared continuum emission, 
and we will use this corrected 
far-infrared continuum in the following discussion.
The spatial resolution of Figure 7b is smoothed to that of our
observation.

First, we concentrate on the general Galactic plane.
Figures 7a and 7b show that the spatial correlation
between the \cii\ line emission and the far-infrared continuum emission
is very good for the general Galactic plane;
most compact sources are common for the both maps,
and the distribution of diffuse components of the \cii\ line emission
is similar to that of far-infrared continuum emission.
Figure 11 shows the intensity correlation between the two emissions;
the two emissions correlate well for the
general Galactic plane data, and the average
$I_{\rm [CII]} / I_{\rm FIR}$ ratio is about 0.6 \%.

One more thing to note in Figure 11  
is that the $I_{\rm [CII]} / I_{\rm FIR}$ ratio has a tendency
to decrease as the \cii\ intensity increases;
the $I_{\rm [CII]} / I_{\rm FIR}$ ratio
is about 0.6 \% for the regions with weak \cii\ line emission
($I_{\rm [CII]} < 1 \times 10^{-4}$ \intensity), 
but decreases to 0.4 \% as the \cii\ line emission becomes stronger
($1 \times 10^{-4} < I_{\rm [CII]} < 3 \times 10^{-4}$ \intensity), 
and reaches almost 0.2 \% at the brightest end 
($I_{\rm [CII]} > 3 \times 10^{-4}$ \intensity).

This situation is illustrated in a different way in Figure 7c,
which shows a $I_{\rm [CII]}/I_{\rm FIR}$ ratio map.
In the general Galactic plane, 
the ratio is rather uniform on a large scale.
However, 
we see some compact dips with $I_{\rm [CII]}/I_{\rm FIR} \sim$
0.2 \%. Theses dips correspond to bright, compact \cii\ sources.

Figure 12 shows
the $I_{\rm [CII]} / I_{\rm FIR}$ ratio vs $I_{60} / I_{100}$ ratio.
Generally speaking, bright \cii\ sources
have higher  $I_{60} / I_{100}$ ratios, which mean
higher UV flux densities (\cite{hol91}).
Figure 12 shows that, as the $I_{60} / I_{100}$ ratio increases,
the $I_{\rm [CII]} / I_{\rm FIR}$ ratio decreases,
i.e. the $I_{\rm [CII]} / I_{\rm FIR}$ ratio
is relatively small toward the regions with high UV flux density.

This tendency can be interpreted qualitatively
on the basis of PDR models
(e.g., \cite{tie85}; \cite{hol91}). 
These model calculations show that,
in high-density regions with high FUV flux density, 
the \cii\ line intensity is rather
saturated due to its lower excitation energy and lower critical
density, and other lines, such as the [\ion{O}{1}] 63 \micron\ 
($^{3}P_{1} \rightarrow$ $^{3}P_{2}$) line, become the dominant coolants. 
Moreover, since dust grains become
positively charged in the regions with high
UV flux density, the total gas heating
efficiency, which is mainly due to the dust photoelectric heating,
also decreases (\cite{hol91}). 
The combination of these two effects decreases the
$I_{\rm [CII]} / I_{\rm FIR}$ ratio toward the sources with high
$I_{60} / I_{100}$ ratios. 

On the other hand, the average $I_{\rm [CII]} / I_{\rm FIR}$ ratio
is high and has large scatter
for the regions with low $I_{60} / I_{100}$ ratios.
The high ratio is attributed to the high gas heating efficiency 
and the dominance of the \cii\ line for gas cooling.
The large scatter is probably due to wide range of physical
conditions in these sources.

These regions with high $I_{\rm [CII]} / I_{\rm FIR}$ ratios
correspond to general diffuse sources.
Thus, in the \cii\ line intensity map, 
the diffuse component is relatively bright.
In other words, the intensity contrast of compact sources 
to diffuse components is relatively small in the \cii\ line map
compared to that
of radio continuum maps and far-infrared continuum maps.

The FIRAS observation showed a good correlation between
$I_{\rm [CII]}$ and $I_{\rm FIR'}$, and Bennet et al. (1994)
obtained a simple power law of $I_{\rm [CII]} \propto 
I_{\rm FIR'}^{0.95}$ ($I_{\rm FIR'}$ is
defined as the continuum intensity at the wavelength of the
\cii\ line). On the other hand, 
Figure 11 shows that we cannot fit our observed data
with a simple power law; 
bright-\cii\ regions have different correlation 
from that of weak-\cii\ regions.
For the regions with $I_{\rm [CII]} < 2 \times 10^{-4}$ \intensity,
we get a correlation of $I_{\rm [CII]} \propto I_{\rm FIR}^{0.83}$,
which is consistent with the FIRAS result.
But the correlation becomes much shallower
($I_{\rm [CII]} \propto I_{\rm FIR}^{0.49}$)
for the regions with $I_{\rm [CII]} \geq 2 \times 10^{-4}$ \intensity.
Since the beam size of FIRAS (7\arcdeg) 
is much larger than typical sizes of active star-forming regions
with bright \cii\ line emission, 
we suppose that the shallower correlation between
$I_{\rm [CII]}$ and  $I_{\rm FIR}$ for star-forming regions
is smeared out in the FIRAS measurement.

\subsection{Galactic Center}

Next, we compare \cii\ line and far-infrared continuum
emissions from the Galactic center.
The spatial correlation between the \cii\ 
emission and the far-infrared continuum is worse than that for the
general Galactic plane (Fig.7a and 7b). 
Moreover, Figure 7c shows that the Galactic center shows a clear, large
dip in the $I_{\rm [CII]} / I_{\rm FIR}$ map.  Figure 11
shows this situation in a different way; 
the intensity correlation between the \cii\ line emission
and the far-infrared continuum emission in the Galactic center region 
is different from that in the general Galactic plane.
As already noted by Nakagawa et al. (1995a), 
the $I_{\rm [CII]} / I_{\rm FIR}$ ratio is about 0.2 \%
in the Galactic center region, 
and is systematically lower than that in the Galactic plane.
This $I_{\rm [CII]} / I_{\rm FIR}$ ratio has no correlation
with the $I_{60} / I_{100}$ ratio (Fig.12), which is again
different from the tendency in the Galactic plane (see above).

Nakagawa et al. (1995a) 
argued that the low $I_{\rm [CII]} / I_{\rm FIR}$
indicates that the abundance of the C$^{+}$ ions
themselves is low in the Galactic center. 
They attributed this low abundance  mainly
to soft UV radiation with fewer C-ionizing photons in the Galactic center.
They also suggested that, together with the molecular
self-shielding due to the high gas density,
this can effectively heat molecular gas on a large scale 
in the Galactic center.
On the basis of the relative weakness of the \cii\ line emission 
to the CO and [\ion{C}{1}] lines measured by FIRAS,
Bennett et al. (1994) also suggested that there is an 
unusually large amount of 
well-shielded dense neutral gas in the Galactic center direction.
Characteristics of the interstellar gas in the Galactic center
will be discussed in a later paper.

\subsection{Velocity Map}

Figure 13 shows the \cii\ line central velocity map, which is
the average of SFP-forth and SFP-back data sets (see $\S$ 3.3), 
together with intensity contours of the \cii\ line emission.
The bin size for the velocity data is 12\arcmin, 
and we show the velocity data only in the bins where
the velocities were determined successfully both in the SFP-forth and
SFP-back data sets.
We determined the central velocity successfully
for most of the regions with $I_{\rm [CII]} > 6 \times 10^{-5}$ \intensity. 

Figure 13 shows a general trend that the dominant
velocity component is blue-shifted in the right side 
and is red-shifted in the left side. This is consistent 
with the Galactic rotation pattern derived from other observations,
such as \ion{H}{1} 21 cm line observations (e.g., \cite{bur88}) 
and CO (1-0) line observations (e.g., \cite{dam87}).
Moreover, we also see some velocity patterns
which do not match the general trend. The most notable example is the steep
velocity gradient across the Galactic center; \vlsr\ 
is about -50 \kms\ at $l$ = -0\fdg6 and is about 40 \kms\
at $l$ = 0\fdg6. Spatially, the former velocity peak
corresponds to Sgr C, and the latter corresponds to Sgr B$_{2}$.
The velocities are also consistent with the main velocity components
of molecular clouds accompanying these \ion{H}{2} regions
(\cite{bal87}). 
Mizutani et al. (1994) also made spectroscopic observations of the 
\cii\ line emission from the Galactic center at $|l| \leq$ 0\fdg7.
The general trend of their observed velocity structures
are consistent with our observed results. 
Although their observed area is rather limited,
they have much better spatial resolution
(3\farcm7) than ours, and they thereby 
revealed more detailed structures of the \cii\ line emission
from the Galactic center.

We also see that each discrete source
has its own velocity (see also Table 2), 
which is different from that
of the surrounding Galactic plane.
For example, the velocity of M~17
(\vlsr\ = 15 \kms) is quite different from that of the 
Galactic plane (\vlsr $\sim$ 50 \kms)
at the same longitude ($l \approx$ 15\arcdeg).
This situation is clearly illustrated also in Figure 4.
The velocity of NGC~6334 (\vlsr\ = $-5$ \kms)
is different from that of the nearby
Galactic plane (\vlsr\ $\approx$ -30 \kms).
One more interesting point of Figure 4 is that the line width is
narrower at NGC~6334 than at the Galactic plane.
This is because there are several velocity components
at the Galactic plane but there is only one
dominant velocity component at each compact source such as NGC 6334.

One more thing we should mention is that, in general, 
clouds at high latitudes show slower velocities
than those at low latitudes. For example, at $l = 17\arcdeg$,
\vlsr\ $\approx 50$ \kms\ at $b$ = 0\arcdeg, but \vlsr\
$\approx 35$ \kms\ at $b = \pm$ 1\arcdeg. This is probably because
the contribution of local components, which generally show slower 
velocities, is more important at higher latitudes than at lower
latitudes. 

\subsection{Longitude-Velocity Map}

Figure 14a shows a longitude-velocity map of the \cii\ emission,
which is obtained by binning the intensities
within $|b| \leq$ 1\arcdeg\
in the longitude-velocity plane with a grid size of
6\arcmin\ in the longitude and 5 \kms\ in the velocity.
Please note that, 
since we obtain only the central velocity and cannot resolve
various velocity components at each line of sight,
this map is not a real longitude-velocity map
but a longitude-``central velocity'' map.

We can see the differential rotation pattern
of our Galaxy very clearly. 
For comparison, we also show a position-velocity 
map of the CO (1-0) line emission (\cite{dam87}) in Figure 14b. 
Although the
instrumental velocity resolution of the BICE system ($\Delta v = 175$ \kms)
is much worse than that of the CO observation ($\Delta v =$
1.3 \kms; \cite{dam87}), the main pattern of the two diagrams
are very similar. This result indicates that the
``fast spectral scanning method'' is very efficient to observe not only line
intensities but also line velocities on a large scale.

However, there are some differences between the two position-velocity 
diagrams. First, we see some wavy structures in the \cii\ diagram.
These wavy
structures correspond to the positions near
bright, compact sources. 
For example, the wavy structure near $l$ = 15\arcdeg\ corresponds
to M~17. As was discussed
above, compact sources generally have different velocities than those
of the surrounding Galactic plane. When there is no bright source at
a certain
longitude, the emission is dominated by that from the Galactic plane,
and the central velocity is that of the Galactic plane. However, if
there is a bright source, the source dominates the \cii\ emission
in the averaged data, and
the central velocity is dragged from that of the Galactic plane
to that of the bright source. Thus the wavy
structures are formed in the position-velocity map 
of the \cii\ line emission (Fig.14a).

Second, the \cii\ position-velocity map shows only 
a main structure and is simpler than the CO position-velocity map.
The CO map shows many 
sub features, such as high velocity components around the Galactic
center, and low velocity components (\vlsr\ $\approx$ 0 \kms) at 
most of the longitudes. 
In the ``fast spectral scanning method'', only one velocity component
is assigned at each observed point, and we cannot resolve various
velocity components along each line of sight.
This procedure makes our \cii\ position-velocity map 
simpler than the CO position-velocity map.

% ---------------------------------------------------------------
\section{CONCLUSION}

(1) We present results of our survey observations
of the \cii\ 158 \micron\ line emission from the Galactic plane.
Our survey covers a wide area
(350\arcdeg\ $\lesssim l \lesssim$ 25\arcdeg, 
$|b| \lesssim$ 3\arcdeg) with a spatial
resolution (15\arcmin) comparable to those of Galactic 
plane surveys at other wavelengths.
 
(2) Strong \cii\ line emission was detected at 
almost all the areas we observed.
Both compact sources and diffuse components were
detected. 

(3) The spatial correlation between the \cii\ line emission
and far-infrared continuum emission is generally good.
However the diffuse component is more eminent
in the \cii\ line emission than in far-infrared continuum emission,
and bright, compact sources show relatively
small $I_{\rm [CII]} / I_{\rm FIR}$ ratios.

(4) The FWHM of the latitudinal profile of the \cii\ 
line (1\fdg32) is slightly wider than
that of far-infrared continuum (1\fdg18).
Both of them are wider than that of the CO line (1\fdg07),
and are much narrower than that of \ion{H}{1} (1\fdg96) line.

(5) The $I_{\rm [CII]} / I_{\rm FIR}$ ratio
is systematically lower in the Galactic center region
on a large scale than that in the Galactic plane
as noted by Nakagawa et al. (1995a)
 
(6) We obtained the the central velocity of the \cii\ line
at each observed point with strong \cii\ line emission
with statistical errors as small as $\pm 6$ \kms.
The results clearly show the differential
rotation of the Galactic disk and also violent velocity 
fields around the Galactic center. 

(7) We employed a new observing method called 
the ``fast spectral scanning''. 
The method was proven to be very efficient for
obtaining not only line intensities but also line velocities
on a large scale.

% ---------------------------------------------------------------
\acknowledgments
We are indebted to the staff of the National Scientific Balloon
Facility, Palestine, Texas (flight numbers 1499p and 1501p). 
This US-Japan collaborative project became
possible through the efforts of L. J. Caroff and M. D. Bicay at NASA
headquarters. We are also grateful to N. Hiromoto for providing us
with far-infrared detectors, to N. Yajima and M. Narita for their
efforts in developing the BICE system, and to H. Murakami for
inspiring discussions. This work was supported  by grants-in-aid from
Ministry of Education, Science, and Culture in Japan, and by NASA.

\clearpage
% ---------------------------------------------------------------

%--------------------------------------------------------------
\clearpage

\figcaption[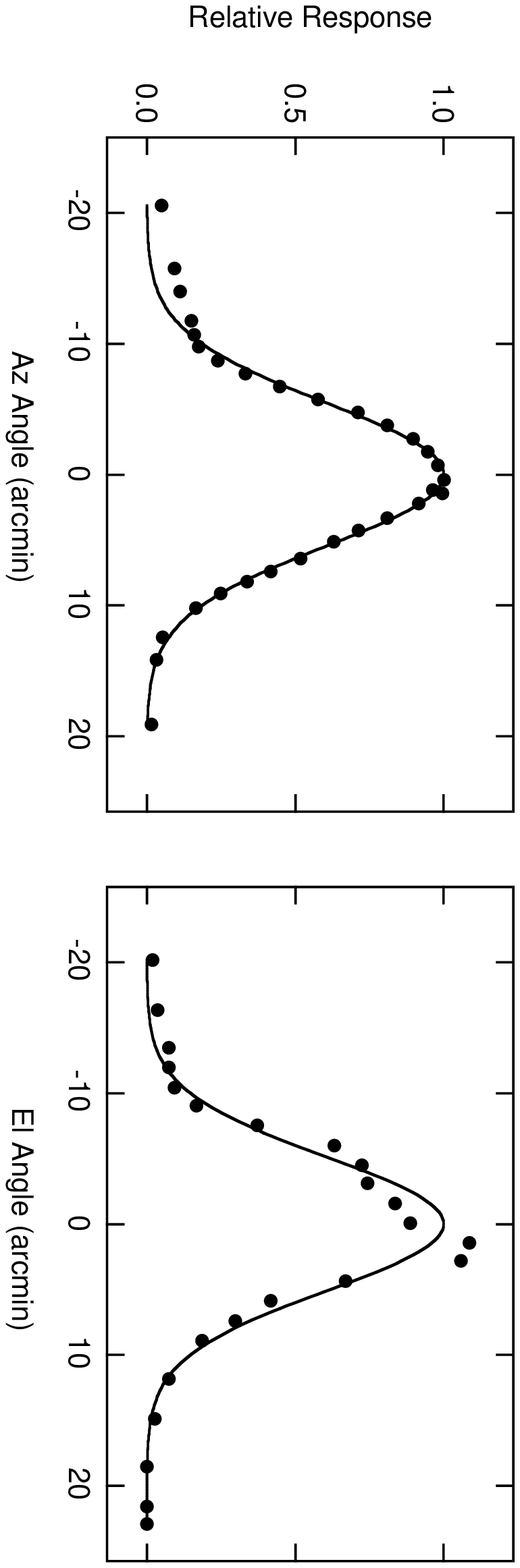]{Beam profiles measured in the laboratory.
	Filled circles show measured data points and curves show
	fitted Gaussian profiles with FWHM of $12\farcm4$.} 

\figcaption[fig2.ps]{The observed area is shown by shades;
	the dark shade indicates the area where the observed
	data are valid, and 
	the light shade shows the area which we used to 
	determine foreground radiation.
	For comparison, far-infrared continuum flux (IRAS, see text) is
	shown by contours; their levels are
	0.3, 0.5, 1, 2, 4, 8, 15, 30, and 60 $\times
	10^{-2}$ erg s$^{-1}$ cm$^{-2}$ sr$^{-1}$.}

\figcaption[fig3.ps]{Scan paths.}

\figcaption[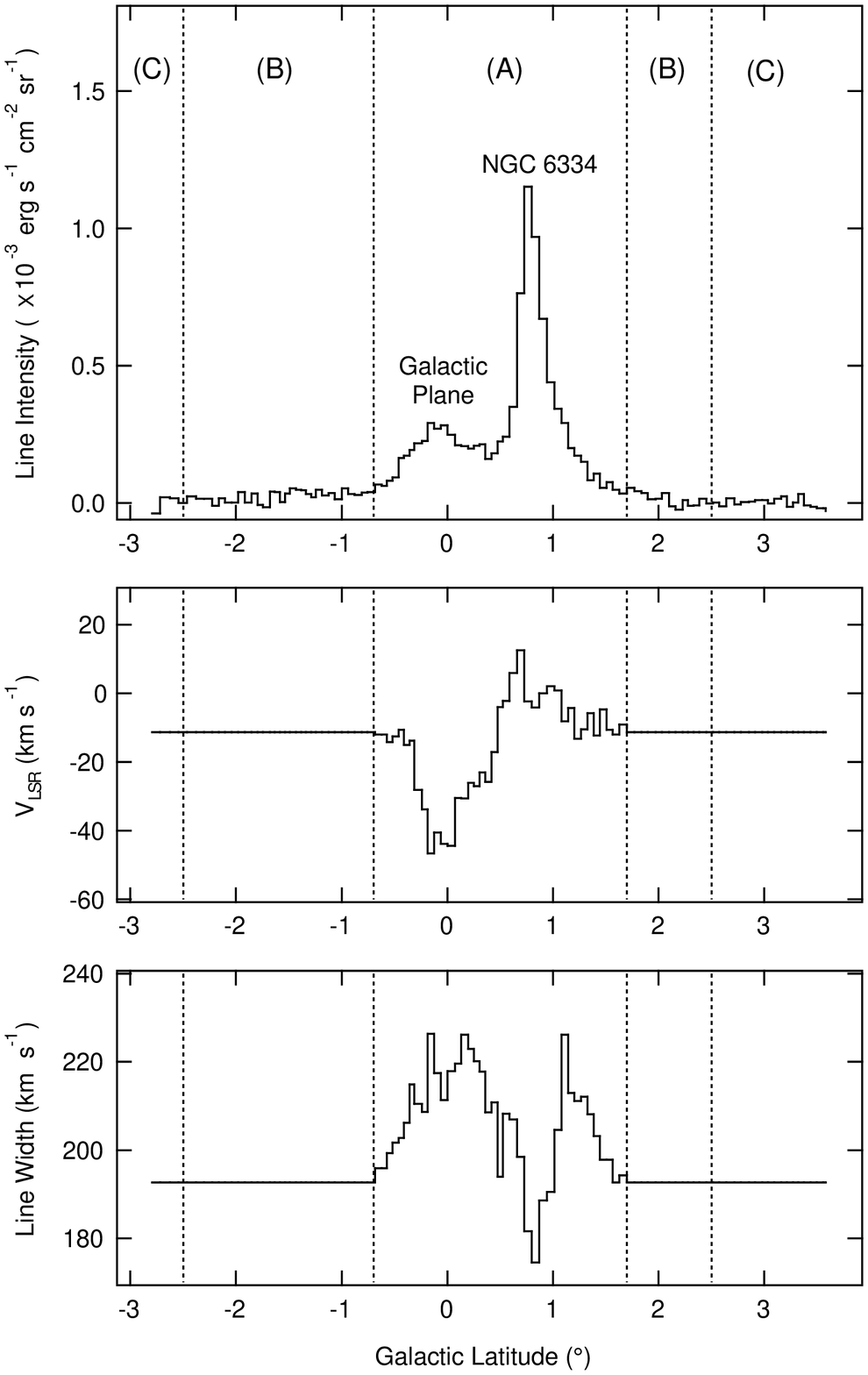]{Example of the data analysis for one spatial scan.
	This scan started at $(l, b) = (-4\fdg5, -2\fdg8)$
	and ended at $(l, b) = (-11\fdg5, -3\fdg6)$.
	The horizontal axis is the Galactic latitude.
	Figure 4a shows the line intensity, Figure 4b shows \vlsr\
	derived from the line central position, and Figure 4c shows
	the line width.
	In the region (A), where the [C II] line intensity is strong, 
	the width, velocity, and intensity of the line
	were derived at each observed point. 
	In the regions (B), where [C II] line intensity is weak, 
	only the line intensity was derived independently at each observed
	point.
	Flat lines in Figure 4b and Figure 4c indicate that
	both the width and the velocity of the line were assumed
	to be constant in one spatial scan (see text).
	The regions (C) are the areas where we determined 
	foreground radiation.}

\figcaption[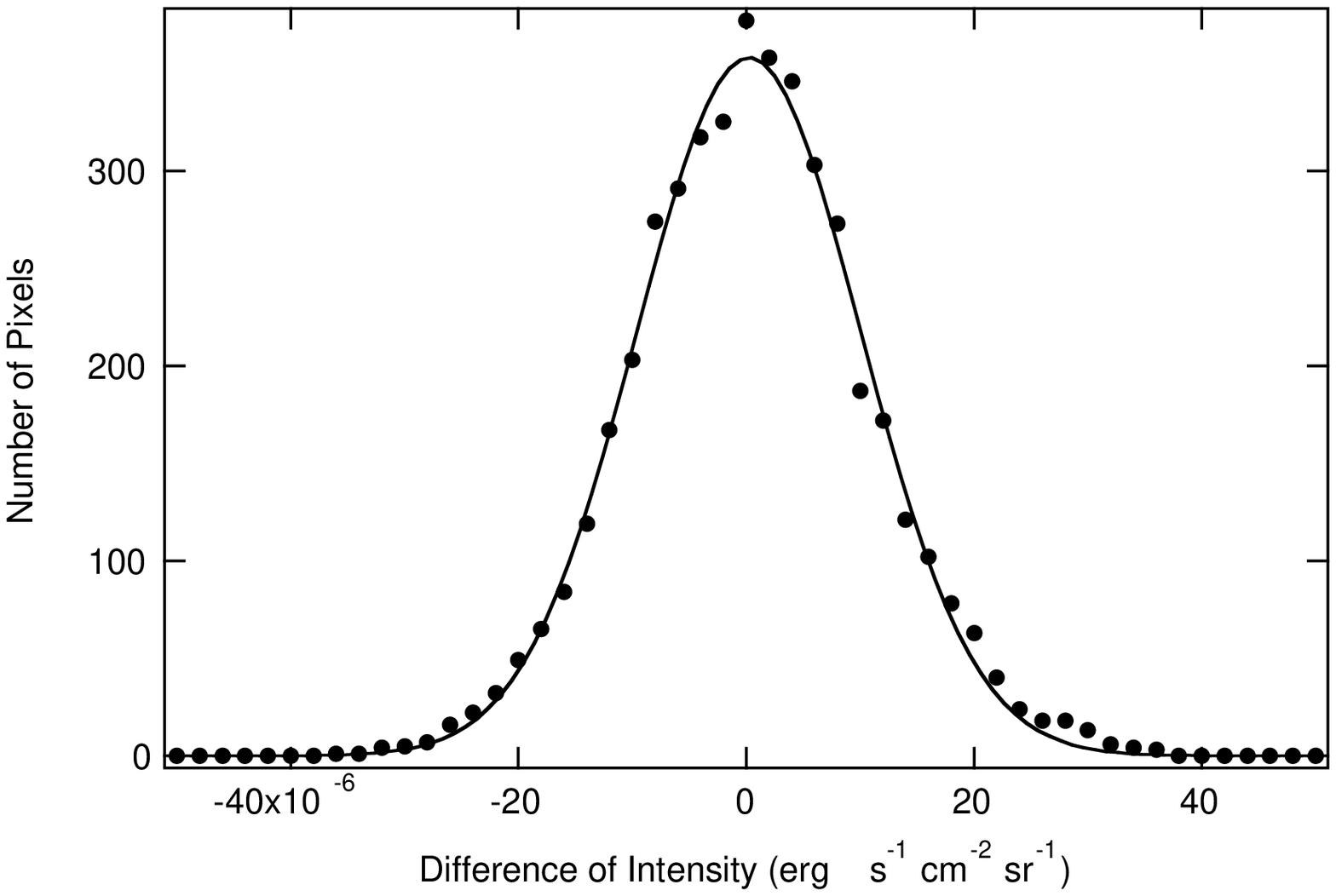]{The difference of intensities between the SFP-forth
	data set and the SFP-back data set. Each data set was binned
	into 3\arcmin\ grids and smoothed to a spatial resolution
	of 15\arcmin\ (FWHM). The solid curve shows a fitted Gaussian
	profile with $\sigma = 1.4 \times  10^{-5}$ \intensity.}

\figcaption[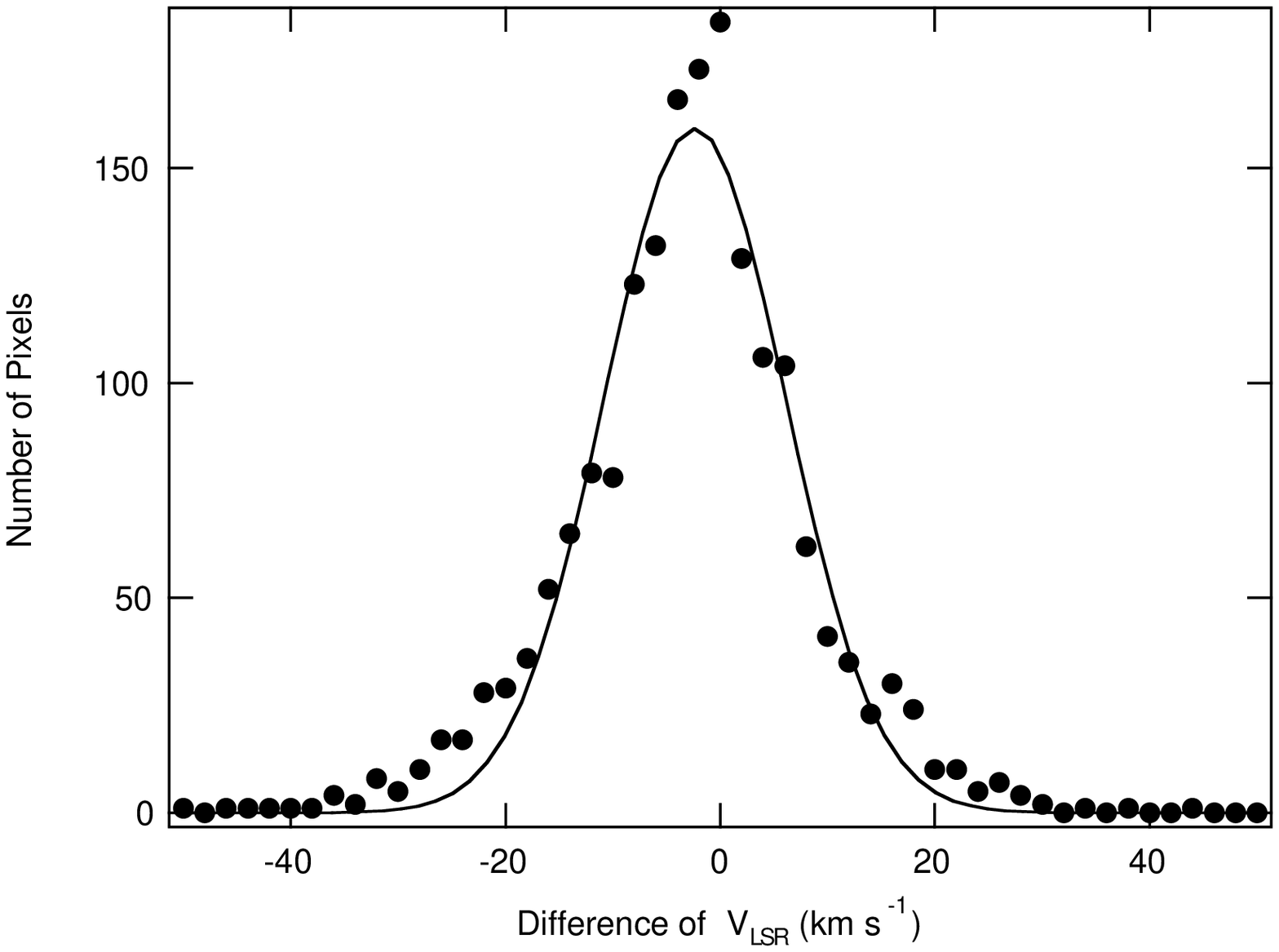]{The difference of velocities between the SFP-forth
	data set and the SFP-back data set. Each data set was binned
	into 12\arcmin\ grids, and no smoothing was made.
	This figure contains only the data point in the areas
	which correspond to the region (A) in Figure 4.
	The solid curve shows a fitted Gaussian
	profile with $\sigma$ = 12 km s$^{-1}$.}

\figcaption[fig7.ps]{(a) Far-infrared [C II] line intensity 
	map ($I_{\rm [CII]}$) obtained 
	by BICE. (b) Far-infrared continuum map ($I_{\rm FIR}$) 
	obtained from IRAS 60 and 100 \micron\ maps.
	Offsets are subtracted (see the text). (c) The ratio of the
	far-infrared [C II] line emission to the far-infrared
	continuum emission ($I_{\rm [CII]}/I_{\rm FIR}$). The
	spatial resolutions of the three maps are 15\arcmin.}

\figcaption[fig8.ps]{Far-infrared [C II] line intensity contour map obtained 
	by BICE with the spatial resolution of 15\arcmin. Contour
	Levels are 0.3, 0.6, 1, 1.5, 2, 3, 4, 6, and 9 $\times 10^{-4}$
	\intensity. The shade shows the observed area.
	Representative bright sources are labelled.}

\figcaption[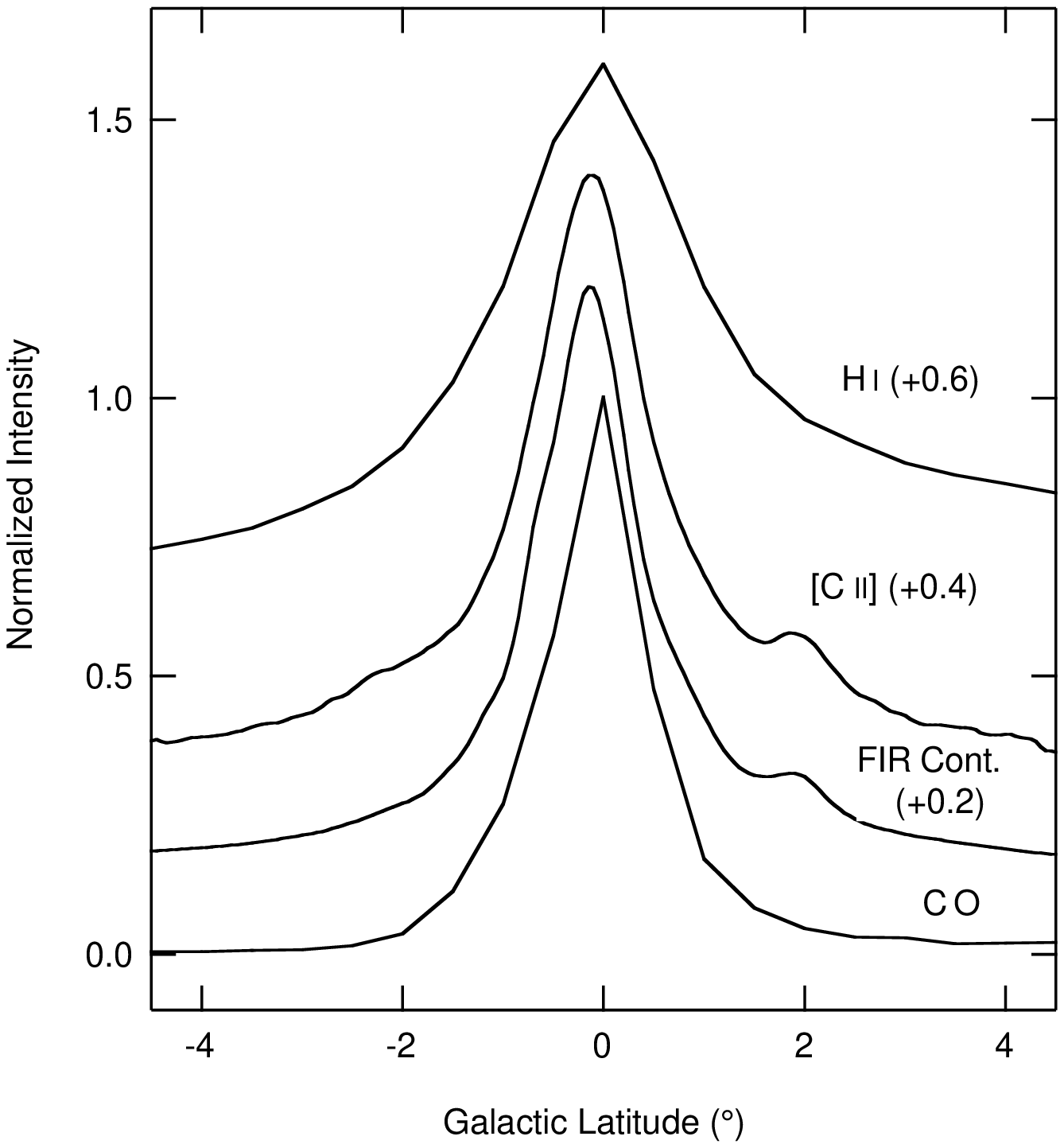]{Latitudinal profile of the [C II] 
	line emission together with those of H~I line 
	(Hartman \& Burton 1997), far-infrared continuum (IRAS), 
	and CO 1-0 line (Dame et al. 1987). The profiles are averaged 
	at 5\arcdeg\ $\leq l \leq$ 25\arcdeg.
	The spatial resolutions of the [C II] line emission and 
	far-infrared continuum profiles are 15\arcmin, and
	bin sizes of H~I and CO profiles are 30\arcmin.
	Peak intensities are normalized, and offsets (values indicated
	in parentheses) are added for clarity.}

\figcaption[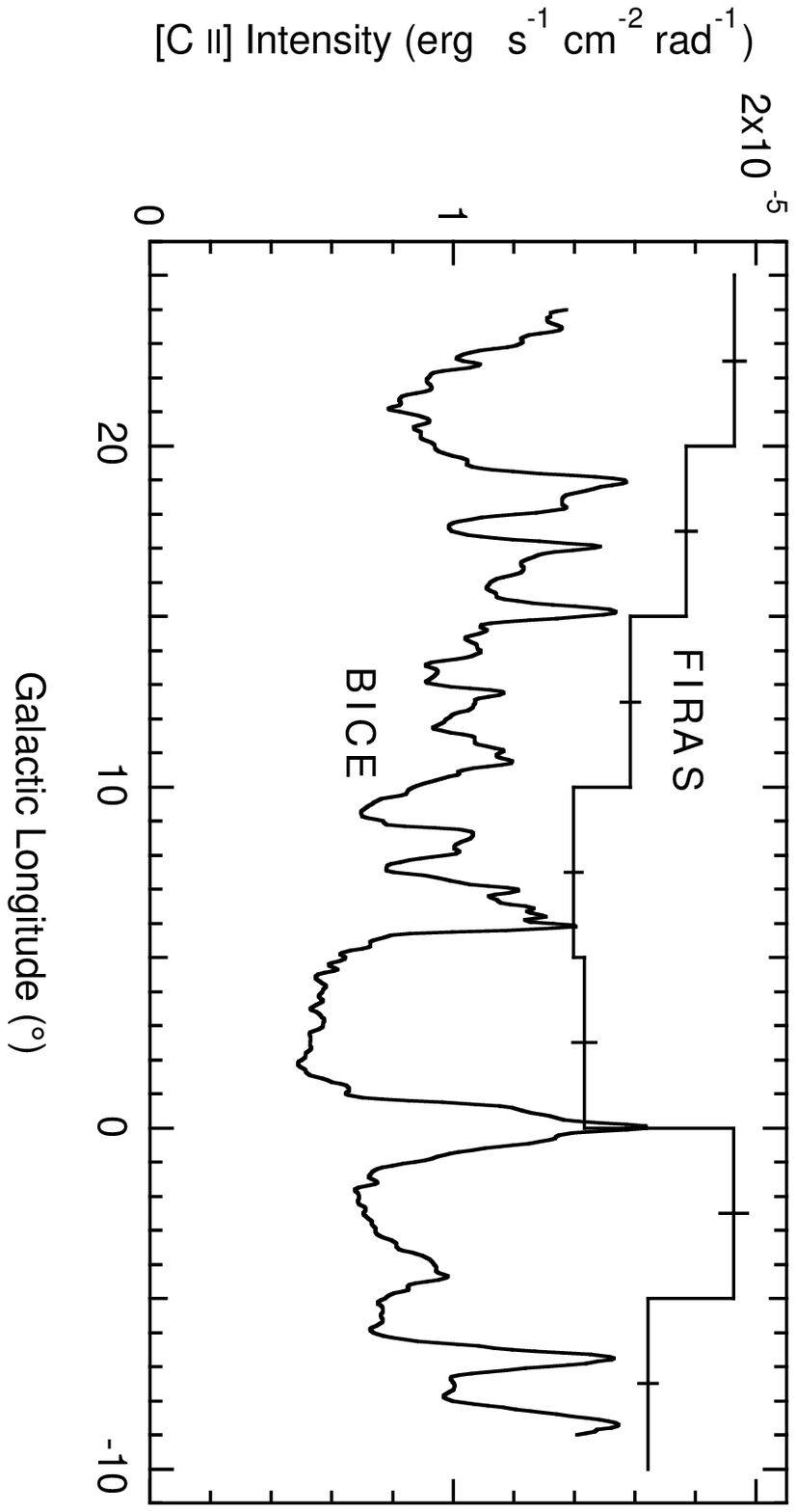]{Longitudinal distribution of the [C II] line
	emission observed by BICE ($|b| \leq$ 3\arcdeg) 
	together with that by FIRAS ($|b| \leq$ 5\arcdeg) 
	(Bennett et al. 1994). We added an uniform offset of 2.3 $\times
	10^{-5}$ erg s$^{-1}$ cm$^{-2}$ sr$^{-1}$ 
	to the BICE result (see text).}

\figcaption[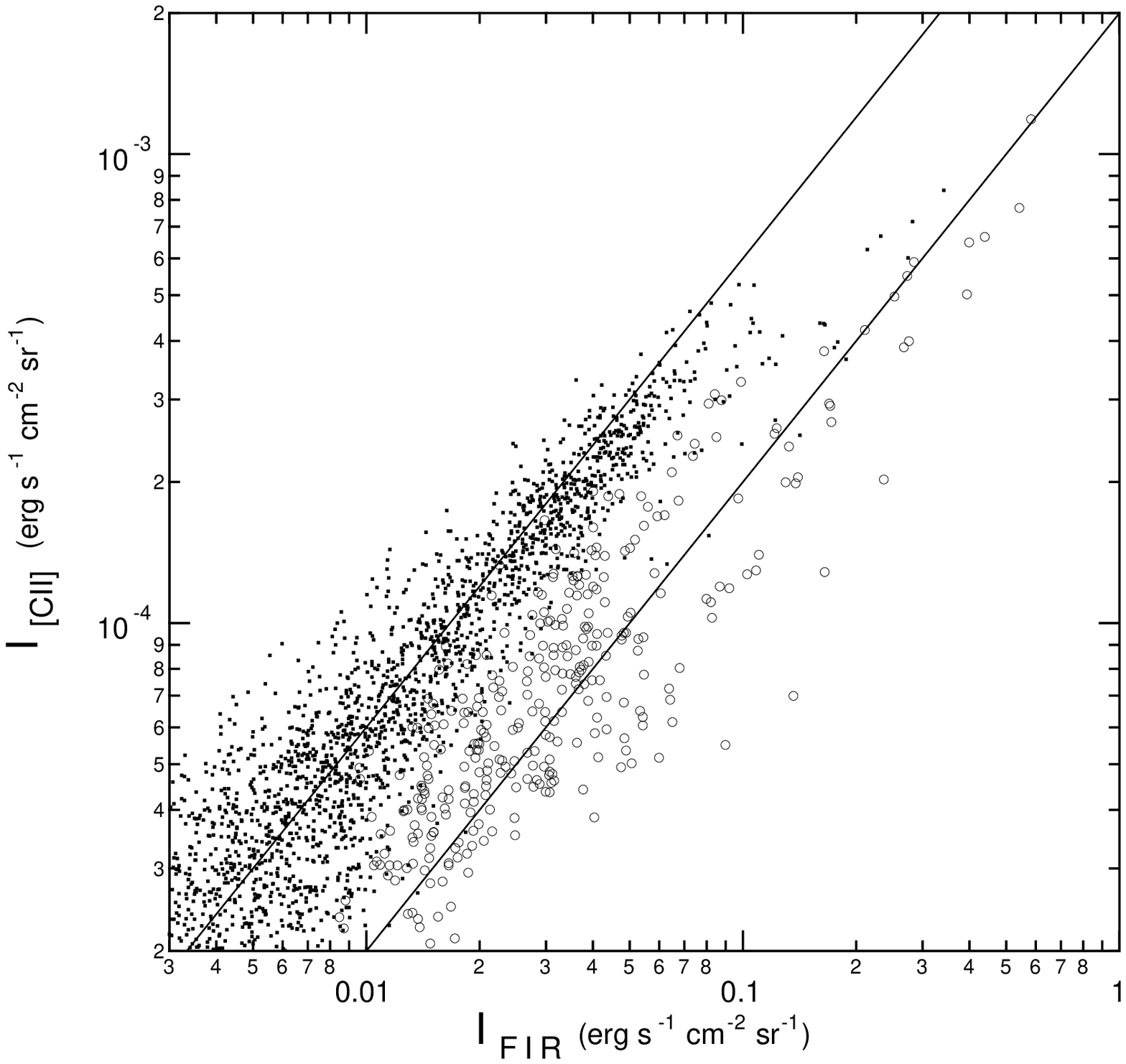]{Intensity correlation between the far-infrared [C II]
	line emission and the far-infrared continuum emission. Each
	point shows a data for a 15\arcmin\ pixel. Open circles are for
	the Galactic center ($-4\arcdeg \leq l \leq 5\arcdeg, |b| \leq 
	1\arcdeg$), and small dots are for the Galactic plane.
	The upper and lower lines show $I_{\rm [CII]}/I_{\rm FIR} = 6 \times
	10^{-3}$ and $2 \times 10^{-3}$, respectively.}

\figcaption[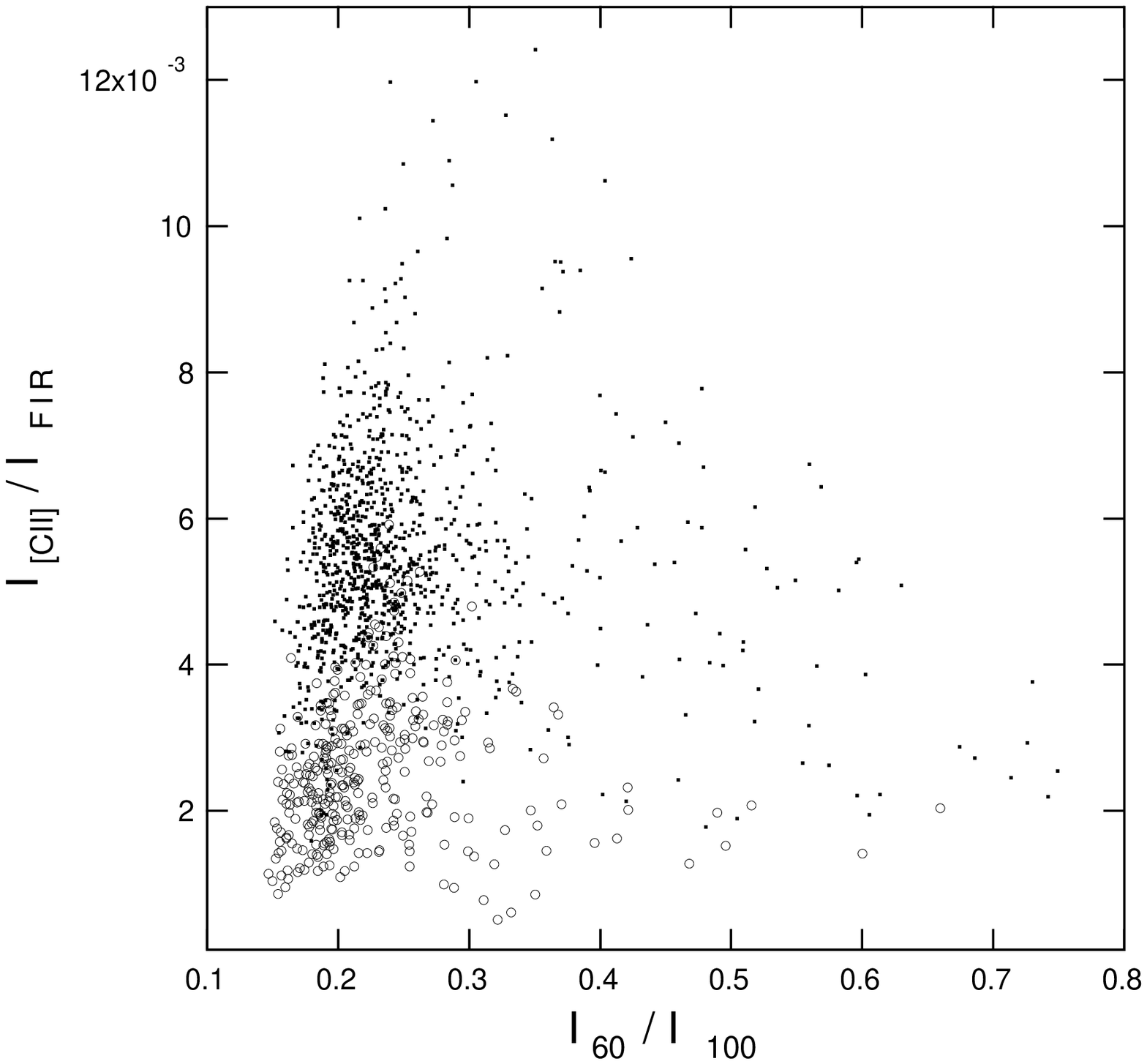]{Relation between the $I_{\rm [CII]}/I_{\rm
	FIR}$ ratio and the far-infrared color
	$I_{60}/I_{100}$. Each
	point shows data for a 15\arcmin\ pixel
	with $I_{\rm FIR} \geq 1 \times 10^{-2}$ \intensity.
	Open circles are for
	the Galactic center ($-4\arcdeg \leq l \leq 5\arcdeg, |b| \leq 
	1\arcdeg$), and small dots are for Galactic plane.}

\figcaption[fig13.ps]{Velocity map of the far-infrared [C II] line
	emission with a pixel size of 12\arcmin.
	The intensity of the [C II] is also shown by contours
	with the levels of 1, 2, 4, and 8 $\times 10^{-4}$
	\intensity.}

\figcaption[fig14.ps]{(a) Longitude-velocity map of the far-infrared
	[C II] line emission. This is obtained by averaging the velocity data
	weighted by intensities at each longitude within $|b| \leq$ 1\arcdeg.
	The results are binned by 6\arcmin\ in the longitude and 5 \kms\
	in the velocity. (b) [C II] longitude-velocity map (contours,
	this work) superposed on a CO longitude-velocity map (false colors, 
	Dame et al. 1987). The resolution of the CO data is 30 \arcmin\ in
	longitude and 1.3 \kms\ in velocity.}
%--------------------------------------------------------------
\clearpage

\begin{deluxetable}{ll}
\tablewidth{0pt}
\tablecolumns{2}
\tablecaption{Observation Parameters}
\tablehead{\colhead{parameter} & \colhead{value}}
\startdata
Target Line             & [C~{\small II}], 
                                $^{2}P_{3/2} \rightarrow$
                                $^{2}P_{1/2}$, 157.7409 \micron \\
Spatial Resolution      & 12$.\mkern-4mu^\prime4$ (FWHM) (beam size) \\
                        & 15$^\prime$ (FWHM) (final map) \\
NEP$_{\rm sys}$         & $6 \times 10^{-16}$ W Hz$^{-1/2}$ \\
Detection Limit    	& $2 \times 10^{-5}$ 
                        ergs s$^{-1}$ cm$^{-2}$ ster$^{-1}$ (3$\sigma$) \\
Velocity Resolution     & 175 km s$^{-1}$ (FWHM) \\
Velocity Determination  & $\pm$6  km s$^{-1}$ (1$\sigma$) \\
Intensity Calibration	& \cii\ line map of M~17 by Matsuhara et al. (1989)\\
Intensity Calibration Uncertainty	& $\pm$ 35 \% \\
Observed Region		& 350\arcdeg $\lesssim l \lesssim$ 25\arcdeg,
			$|b| \lesssim$ 3\arcdeg\\
\enddata
\end{deluxetable}

%--------------------------------------------------------------
\clearpage

\begin{deluxetable}{rrrrcccrl}
\tablewidth{0pt}
\tablecolumns{9}
\tablecaption{Bright Peaks}

\tablehead{
\colhead{No}&
\colhead{$l$} 		& \colhead{$b$} & 
\colhead{peak} 		& \colhead{base} &	
\colhead{$\Delta l$} 	& \colhead{$\Delta b$} & 
\colhead{$v_{\rm LSR}$}	& \colhead{identification}
\\
\colhead{} &
\colhead{(\arcdeg)} 	& \colhead{(\arcdeg)} & 
\colhead{intensity\tablenotemark{a}} 	& 
\colhead{level\tablenotemark{a}} &	
\colhead{(\arcmin)} 	& \colhead{(\arcmin)} & 
\colhead{(km s$^{-1}$)}	& \colhead{}
}

\startdata
1 &
351.30 & 0.70  & 9.07\phm{AA} & 1.20 & 29 & 23 & -5\phm{AA} & NGC~6334\nl
2 &
353.20 & 0.70  & 6.10\phm{AA} & 1.00 & 26 & 37 &  5\phm{AA} & NGC~6357\nl
3 &
353.55 & 0.00  & 3.39\phm{TA} & 1.83 & 16 & 15 & -34\phm{PA} & \nl
4 &
0.05  & 0.00 & 12.02\phm{AA} & 6.90 & 15 & 19 & -4\phm{AA} & Sgr~A\nl
5 &
5.90  & -0.40 & 3.83\phm{AA} & 2.20 & 17 & 18 & 10\phm{AA} & \nl
6 &
5.95  & -1.15 & 4.59\phm{AA} & 1.50 & 20 & 31 & 11\phm{AA} & M~8\nl
7 &
6.60  & -0.20 & 3.45\phm{AA} & 2.65 & 15 & 16 & 21\phm{AA} & \nl
8 &
7.05  & -0.20 & 3.37\phm{AA} & 2.65 & 15 & 15 & 16\phm{AA} & M~20\nl
9 &
8.15  & 0.00  & 3.12\phm{AA} & 2.30 & 19 & 15 & 42\phm{AA} & \nl
10 &
8.60  & -0.30 & 3.81\phm{AA} & 2.60 & 24 & 28 & 30\phm{AA} & W~30\nl
11\tablenotemark{b} &
10.40 & 0.00 & 3.24\phm{AA} & \nodata & \nodata & \nodata &  7\phm{AA} & \nl
12 &
10.70 & -0.35 & 3.66\phm{AA} & 2.70 & 17 & 15 &  8\phm{AA} & \nl
13 &
12.80 & -0.20 & 3.91\phm{AA} & 2.10 & 19 & 19 & 27\phm{AA} & \nl
14 &
14.00 & -0.10 & 3.32\phm{AA} & 2.56 & 29 & 15 & 37\phm{AA} & \nl
15\tablenotemark{b} &
14.10 & -0.55 & 3.04\phm{AA} & \nodata & \nodata & \nodata &  30\phm{AA} & \nl
16 &
15.10 & -0.70 & 8.71\phm{AA} & 1.50 & 25 & 20 & 15\phm{AA} & M~17\nl
17 &
16.40 & 0.05  & 3.55\phm{AA} & 2.55 & 17 & 24 & 46\phm{AA} & \nl
18\tablenotemark{b} &
16.45 & -0.20 & 3.16\phm{AA} & \nodata & \nodata & \nodata & 43\phm{AA} \nl
19 &
17.10 & 0.90  & 4.59\phm{AA} & 1.40 & 23 & 34 & 32\phm{AA} & M~16\nl
20 &
18.20 & -0.25 & 4.49\phm{AA} & 2.60 & 16 & 18 & 51\phm{AA} & \nl
21 &
18.15 & 2.00  & 3.51\phm{AA} & 0.53 &    &    & 35\phm{AA} & 
\vspace{-0.4em}
\nl
22 &
18.40 & 2.00  & 3.48\phm{AA} & 0.52 &$\Bigg\}$ 33 & 29 & 35\phm{AA} & 
$\Bigg\}$S~54, W~35
\vspace{-0.75em}
\nl
23 &
18.75 & 2.00  & 3.36\phm{AA} & 0.49 &    &    & 26\phm{AA} & \nl
24 &
19.05 & -0.40 & 4.94\phm{AA} & 2.60 & 22 & 37 & 62\phm{AA} & \nl
25 &
19.60 & -0.10 & 3.16\phm{AA} & 1.95 & 23 & 20 & 60\phm{AA} & \nl
26 &
20.75 & -0.10 & 3.07\phm{AA} & 1.60 & 20 & 18 & 62\phm{AA} & \nl
27\tablenotemark{c} &
22.30 & -0.20 & 3.35\phm{AA} & \nodata & \nodata & \nodata & 83\phm{AA} & \nl
28\tablenotemark{c} &
23.00 & -0.35 & 4.62\phm{AA} & \nodata & \nodata & \nodata & 71\phm{AA} & \nl
29 &
23.80 & 0.10  & 4.11\phm{AA} & 3.15 & 21 & 15 & 91\phm{AA} & \nl 
30\tablenotemark{c} &
24.60 & -0.10 & 5.61\phm{AA} & \nodata & \nodata & \nodata & 88\phm{AA} & \nl
\enddata

\tablenotetext{a}{The unit is $10^{-4}$ \intensity.}
\tablenotetext{b}{These sources are parts of diffuse emission, and
we cannot determine their base levels and source sizes.}
\tablenotetext{c}{These sources are at the edge of scanned areas, and
part of the sources are out of our observed range.
Hence base levels and source sizes cannot be determined precisely. 
The true peaks may be out of our observed area and 
their intensities are also uncertain. The peak intensities listed above
are those within our surveyed areas.}

\end{deluxetable}

%---------------------------------------------

\end{document}